\documentclass[preprint, jcp, aip, a4paper, amsmath, superscriptaddress, floatfix]{revtex4-1}

\usepackage[hidelinks,
    pdfauthor={C. Arnold, L. Inhester, S. Carbajo, R. Welsch, R. Santra},
    pdftitle={ Simulated XUV Photoelectron Spectra of THz-pumped Liquid Water }
]{hyperref}

\usepackage{datetime}

\usepackage[T1]{fontenc}
\usepackage[utf8]{inputenc}
\usepackage{color}

\usepackage[english]{babel}

\usepackage{amsmath}

\usepackage{units}
\usepackage{gensymb}
\usepackage{braket}
\usepackage{booktabs}
\usepackage{graphicx}
\graphicspath{ {/} {./graphics/} {./graphics/plots/} }

\newcommand{\chem}[1]{\ensuremath{\mathrm{#1}}}
\newcommand{\vect}[1]{\ensuremath{\mathbf{#1}}}

\DeclareUnicodeCharacter{00B0}{\degree}

\newcommand*{\incite}[1]{%
  \begingroup
    \romannumeral-`\x % remove space at the beginning of \setcitestyle
    \setcitestyle{numbers}%
    \cite{#1}%
  \endgroup   
}

\begin{document}

\title{ Simulated XUV Photoelectron Spectra of THz-pumped Liquid Water }

\author{Caroline Arnold}
\email{caroline.arnold@cfel.de}
\affiliation{Center for Free-Electron Laser Science, DESY, Notkestrasse 85, 22607 Hamburg, Germany}
\affiliation{Department of Physics, University of Hamburg, Jungiusstrasse 9, 20355 Hamburg, Germany}
\affiliation{The Hamburg Centre for Ultrafast Imaging, Luruper Chaussee 149, 22761 Hamburg, Germany}

\author{Ludger Inhester}
\affiliation{Center for Free-Electron Laser Science, DESY, Notkestrasse 85, 22607 Hamburg, Germany}

\author{Sergio Carbajo}
\affiliation{Stanford University and SLAC National Accelerator Laboratory, 2575 Sand Hill Road, Menlo Park, CA 94025, USA}

\author{Ralph Welsch}
\email{ralph.welsch@cfel.de}
\affiliation{Center for Free-Electron Laser Science, DESY, Notkestrasse 85, 22607 Hamburg, Germany}

\author{Robin Santra}
\affiliation{Center for Free-Electron Laser Science, DESY, Notkestrasse 85, 22607 Hamburg, Germany}
\affiliation{Department of Physics, University of Hamburg, Jungiusstrasse 9, 20355 Hamburg, Germany}
\affiliation{The Hamburg Centre for Ultrafast Imaging, Luruper Chaussee 149, 22761 Hamburg, Germany}

The following article has been accepted for publication by Journal of Chemical Physics. After it is published, it will be found at \url{https://doi.org/10.1063/1.5054272/}.

\begin{abstract}
    Highly intense, sub-picosecond terahertz (THz) pulses can be used to induce ultrafast temperature jumps (T-jumps) in liquid water. A supercritical state of gas-like water with liquid density is established, and the accompanying structural changes are expected to give rise to time-dependent chemical shifts. We investigate the possibility of using extreme ultraviolet (XUV) photoelectron spectroscopy as a probe for ultrafast dynamics induced by sub-picosecond THz pulses of varying intensities and frequencies. To this end, we use \textit{ab initio} methods to calculate photoionization cross sections and photoelectron energies of $\chem{(H_2O)_{20}}$ clusters embedded in an aqueous environment represented by point charges. The cluster geometries are sampled from \textit{ab initio} molecular dynamics simulations modeling the THz-water interactions. We find that the peaks in the valence photoelectron spectrum are shifted by up to $\unit[0.4]{eV}$ after the pump pulse, and that they are broadened with respect to unheated water. The shifts can be connected to structural changes caused by the heating, but due to saturation effects they are not sensitive enough to serve as a thermometer for T-jumped water. 
\end{abstract}

\maketitle

%================================================================================ 
% TEXT
%================================================================================ 

\section{Introduction}

Water is the main solvent on earth and the primary component of living organisms. The ultrafast processes that unfold in water on a femto- to picosecond time scale are thus of direct relevance to many biological and chemical processes, e.g., radiation damage and reaction kinetics \cite{ball17,garrett05}. To investigate these ultrafast chemical reactions, one requires means to trigger them in a controlled and abrupt way. It is therefore desirable to immediately transfer energy into specific degrees of freedom that initiate the sought-after ultrafast processes.  To this end, short pulses that directly target vibrational modes can be used. In the infrared (IR) range intramolecular modes are excited, while in the terahertz (THz) range the intermolecular librations (hindered rotations) and vibrations are pumped. Large amounts of energy can be transferred from ultrashort, high-intensity THz pulses to water through the efficient coupling of THz radiation to water molecules due to the large dipole moment of water. This results in an ultrafast increase in temperature known as a temperature jump (T-jump).

T-jump experiments were originally introduced to measure rates of chemical reactions in solutions on the timescale of microseconds \cite{crooks83}. The advent of femtosecond lasers allowed to investigate the fundamental timescale of chemical dynamics \cite{ma06}. Using IR laser light in resonance with intramolecular O-H modes, a T-jump of few Kelvin on the timescale of picoseconds has been observed in liquid water through transient x-ray absorption \cite{wernet08}. Similar techniques were used to resolve ultrafast changes in the hydrogen bond network \cite{huse09}. Exciting intermolecular modes in the THz range is of special interest, as they are tied directly to the hydrogen bond network that is assumed to play a role in the numerous water anomalies \cite{nilsson15}, as well as the solvation dynamics \cite{continibali14}. These processes can be investigated by 2D Raman-THz spectroscopy \cite{finneran16,finneran17,savolainen13}. However, these types of experiments are very hard to interpret \cite{finneran17}, and thus more direct approaches to investigate the structural dynamics and the role of the hydrogen-bond network in solvation dynamics of THz-pumped water are desirable.

THz-induced T-jumps can also be used to vibrationally excite other inorganic or organic molecules of interest that are solvated in liquid water \cite{mishra15}. Here, the THz pulse does not directly couple to vibrational modes of the solvated molecule, but energy is redistributed through the excitation of solvent modes. This allows to trigger ultrafast molecular dynamics in these molecules that are not accessible by current techniques in ultrafast, time-resolved experiments that typically trigger reactions by direct, electronic excitation with an ultrashort pulse. This opens new possibilities to explore in the field of ultrafast chemistry.

High-intensity THz pulses of a duration of less than a picosecond have recently been realized experimentally in both lab-based sources \cite{hoffmann11, hafez16} and as part of free-electron laser (FEL) beamlines \cite{stojanovic13, zhang17}.  At FELs like, e.g., FLASH, split THz-XUV beamlines are available \cite{kuntzsch12,tiedtke09}. At FLASH, THz pulses are created by the same electron bunch as the XUV beam, producing aligned and synchronized pulses. This setup is ideally suited for pump-probe experiments, where a THz pulse triggers dynamics in liquid water that are then probed by XUV spectroscopy. The short probe pulse duration of \unit[50 to 100]{fs} enables the time-resolution of ultrafast processes. An even better time resolution is conceivable considering lab-based probe-pulse sources \cite{huppert15, jordan18}.

Theoretical descriptions of the sub-picosecond, THz-induced T-jump in liquid water have been given employing both classical force fields \cite{mishra18, yang17,mishra16,huang16a} and \textit{ab initio} molecular dynamics (AIMD) \cite{mishra13,mishra18}. The maximum energy transfer is reached with pump pulse frequencies between 14 and 17 THz, where rotations of water molecules are excited \cite{mishra18,huang16a}. Ultrafast T-jumps of more than $\unit[1000]{K}$ within tens of femtoseconds are predicted within currently accessible THz intensity levels \cite{mishra18}. The instantaneous temperature can be calculated directly from the molecular dynamics (MD) trajectories of THz-pumped water and was shown to be independent of the simulation box size \cite{huang16a}. It can be connected to structural changes of the pumped water. Observing these structural changes in experiments with time resolution of less than 100 fs can be used to evaluate the efficiency of the THz-induced T-jump. 

Measuring ultrafast T-jumps experimentally is challenging. Generally, in ultrafast pump-probe experiments, structural changes may be probed employing two different strategies. First, by x-ray diffraction techniques the rearrangement of nuclear geometries can be directly investigated. The structure factor, related by Fourier transform to the radial distribution function (RDF), is an indicator for the order maintained in a liquid. Especially in water, vanishing peaks in the RDF have been used as indicators of the rupture of hydrogen bonds in supercritical conditions \cite{chialvo94,hoffmann97}. 
%
% TODO krack02, hura03
%
%With the advent of coherent x-ray sources, x-ray photon correlation spectroscopy (XPCS) can be used to study non-equilibrium dynamics of liquid water \cite{perakis18, shpyrko14}. 
%
Second, changes in the molecular geometries have impact on the electronic structure \cite{winter07,wernet11, zalden18}. The supercritical state of water that can be created by sub-picosecond THz pump pulses, where the T-jump occurs isochorically, will be accompanied by changes in the valence electron levels that are susceptible to changes in the hydrogen bond network \cite{wernet05,guo02,chen10,wang16a,nishizawa10}. In the context of IR-induced T-jumps in water, x-ray absorption has been used as a probe \cite{wernet08}. With photon energies close to the oxygen K-edge, low-lying unoccupied molecular orbitals are investigated. The complementary tool to study shifts in the occupied outer and inner valence molecular orbitals is x-ray emission spectroscopy \cite{yin17} or XUV photoelectron spectroscopy \cite{link09a}. 

In this manuscript, we will explore the potential of photoelectron spectroscopy as a probe for the structural changes induced by an ultrafast T-jump in liquid water following the excitation by high-intensity, sub-picosecond THz pulses of various peak intensities and frequencies. In Sec.~\ref{sect:structural-changes}, we analyze the structural changes that are induced by sub-picosecond, high-intensity THz pulses based on AIMD trajectories \cite{mishra18}. We then describe the theoretical framework for the calculation of photoionization cross sections from liquid water in Sec.~\ref{sect:photoionization-theory}. The simulated photoelectron spectrum of liquid water is discussed in Sec.~\ref{sect:results-unheated} and characteristic changes in the photoelectron spectrum following a high-intensity THz pump pulse are investigated in Sec.~\ref{sect:results-single-pulse}. Finally, in Sec.~\ref{sect:outlook} we summarize our results and open up connections to possible experiments. 

% Structural changes induced by THz pumping
\section{Structural Changes Induced by THz-Pumping}
\label{sect:structural-changes}

\subsection{ \textit{Ab initio} Molecular Dynamics Trajectories}
\label{sect:aimd-trajectories}

The analysis in this work is based on \textit{ab initio} molecular dynamics (AIMD) trajectories of THz-pumped water that were the subject of previous work\cite{mishra18}. The trajectories were generated with the CP2K molecular dynamics package \cite{kuhne07}, employing the Perdew-Burker-Ernzerhof (PBE) functional together with the Geodecker-Teter-Hutter (GTH) pseudopotential, as well as the TZV2P basis set. The non-empirical PBE functional was chosen by Mishra \textit{et al.} as it reproduces electronic polarizability well. This property determines the coupling to the THz pump pulse, and a good description is thus imperative to study THz-induced T-jumps. For more details, we refer to Refs.~\incite{mishra13, mishra18}. Nonetheless, please note that the PBE functional results in water with liquid properties that deviate from bulk water at room temperature \cite{lin12, heyden10}. Mishra \textit{et al.} obtained ten different initial conditions by taking snapshots from an \textit{ab initio} trajectory under NVT conditions, stabilized with the Nose-Hoover thermostat, a temperature of \unit[300]{K} and a fixed density of $\unit[1]{g/cm^3}$. A total number of 128 water molecules was used with periodic boundary conditions for a cubic box of $\unit[16]{\mathring A}$ edge length. Mishra \textit{et al.} sampled ten initial conditions with a time separation of \unit[300]{fs}. Since the the PBE functional underestimates diffusion constants and prolongates equilibration times \cite{lin12}, this resulted in a slightly favored orientation of the water molecules in the initial trajectory snapshots with $\langle\cos\vartheta\rangle = 0.08\dots 0.13$, where $\vartheta$ is the angle between a molecular dipole and the THz field axis. The variance of these angles was $\langle \cos^{2} \vartheta \rangle \approx 1/ 3$ across all initial conditions, which corresponds to the value expected with uniform sampling. The partially incomplete equilibration may cause some short-time non-equilibrium dynamics. As the sub-picosecond, high-intensity THz pulse quickly drives the system far out of equilibrium, this will not affect the main conclusions drawn in this work.

For each pulse intensity and frequency given in Table~\ref{tab:run-parameters}, these initial snapshots were propagated for $\unit[500]{fs}$ including a THz pulse, and for an additional $\unit[1]{ps}$ without any external field. The THz pump pulse was implemented by
\begin{equation}
  \vect E(t) = \varepsilon(t) \vect e_z \cos( 2 \pi \nu t + \varphi),
  \label{eq:thz-pulse}
\end{equation}
where the Gaussian pulse envelope was $\varepsilon(t) = A \exp\left( - t^2 / 2 \sigma^2\right)$, with a width of $\sigma = \tau_\mathrm{FWHM} / 2 \sqrt{\ln 2}, \tau_\mathrm{FWHM}=\unit[150]{fs}$, carrier-envelope phase $\varphi = \pi / 2$, and the central THz frequency $\nu$. The peak field amplitudes considered were $A = \unit[0.274]{V / \mathring{A} }$, corresponding to a peak pulse intensity of $I = \unit[1 \times 10^{12}]{W / cm^2}$, and $A = \unit[0.614]{V / \mathring{A}}$, corresponding to $I = \unit[5 \times 10^{12}]{W / cm^2}$, respectively. 

The THz pump pulse excites intermolecular modes and transfers a large amount of energy to bulk water within less than a picosecond. Prior analysis of the trajectories revealed that the induced T-jump is maximal for $\unit[16]{THz}$\cite{mishra18}. The T-jump scales with intensity obeying a power law, $\Delta \bar T(I) \propto I^{\beta_\nu}$, where a nonlinear regime could be found for $\nu < \unit[19]{THz}$ $(\beta_\nu \approx 1.3 ... 1.4)$, separated from a saturation regime at higher THz frequencies $(\beta_\nu < 1)$. The large amount of energy deposited through the pump pulse led to problems with energy conservation in some of the AIMD trajectories. Therefore, for the high intensity considered, these trajectories had to be excluded from the analysis, reducing the available number of trajectories as summarized in Table~\ref{tab:run-parameters}. The trajectories were generated under isochoric conditions, as the bulk water does not expand considerably during the first \unit[1.2]{ps} following the THz pump pulse \cite{mishra16}. Therefore, all structural changes that can be extracted from the trajectories are related to the excitation of intra- and intermolecular modes. 

% =============================================
% TABLE
\begin{table}[h]
  \caption{ 
    Number of trajectories from different initial conditions, considered for different values of the THz pump pulse intensity $I$ and central frequency $\nu$. 
  }
  \begin{tabular}{ccc}
    \toprule
    $\unit[I]{[W / cm^2]}$   &   $\unit[\nu]{[THz]}$ &   Trajectories\\
    \midrule
%    $1\times 10^{11}$   &   $7, 13, 16, 19, 30$  &   10  &   6   \\
    $1\times 10^{12}$   &   $7, 16, 19, 30$  &   10  \\
    $5\times 10^{12}$   &   $7,10$      &   9     \\
    $5\times 10^{12}$   &   $30$  &   10    \\
    $5\times 10^{12}$   &   $16,19$  &   7     \\
    \bottomrule
  \end{tabular}
  \label{tab:run-parameters}
\end{table}
% =============================================

\subsection{Transformation to Super-Critical State by THz pump}
\label{sect:aimd-quantities}

%================================================================================ 
% BEGIN Figure RDF
\begin{figure*}[h]
    \centering
    \includegraphics{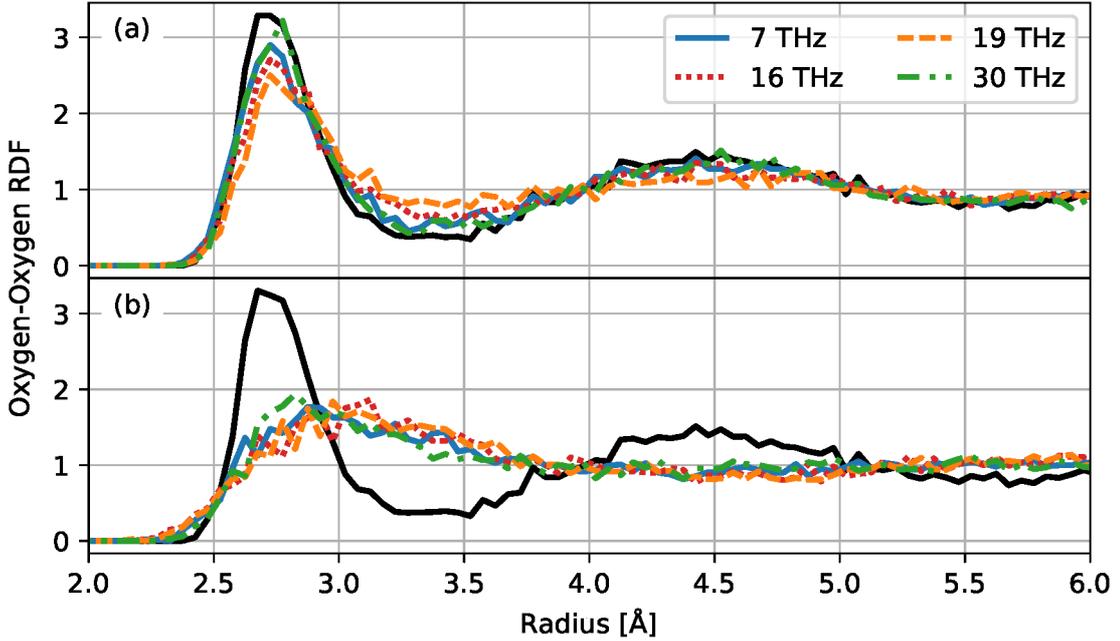}
    \caption{Radial distribution function (RDF) for oxygen pairs for a THz pulse intensity of \textit{(a)}~$\unit[1\times~10^{12}]{W/cm^{2}}$ and \textit{(b)}~$\unit[5\times 10^{12}]{W/cm^2}$, and frequencies of 7, 16, 19, and 30 THz, respectively. The RDF is calculated from trajectory snapshots directly after the end of the pump pulse, $\unit[250]{fs}$ after the pulse maximum, with 128 molecules employing periodic boundary conditions and averaged over up to ten trajectories, cf.~Table~\ref{tab:run-parameters}. The RDF of unheated water, i.e., $t=\unit[-250]{fs}$, is shown as a black solid line. }
    \label{fig:rdf_OO}
\end{figure*}
% END Figure RDF
%================================================================================ 

%
\begin{table}
    \centering
    \caption{Quantities evaluated after the THz pulse at $t=\unit[250]{fs}$: T-jump values $\Delta T$ repeated from Mishra \textit{et al.}~\cite{mishra18} and remaining fraction $f_\mathrm{HB}$ of hydrogen bonds. }
    \label{tab:aimd-quantities}
    \begin{tabular}{cccc}
        \toprule
        $\unit[I]{[W/cm^2]}$   &%
        $\unit[\nu]{[THz]}$    &%
        $\unit[\Delta T]{[K]}$ &%
        $f_\mathrm{HB}$
        \\
        \midrule
        $1\times 10^{12}$ & 7  & 76   & 0.84 \\
        $1\times 10^{12}$ & 19 & 212  & 0.44 \\
        $1\times 10^{12}$ & 30 & 128  & 0.78 \\
        $5\times 10^{12}$ & 7  & 721  & 0.11 \\
        $5\times 10^{12}$ & 19 & 1193 & 0.06 \\
        $5\times 10^{12}$ & 30 & 558  & 0.19 \\
        \bottomrule
    \end{tabular}
\end{table}
% RDF
From the AIMD trajectories, the radial distribution function (RDF) of heated water after the pulse ($t=\unit[250]{fs}$, where $t=\unit[0]{fs}$ refers to the maximum of the pump pulse envelope) is calculated for different pump pulse frequencies, see Fig.~\ref{fig:rdf_OO}. At $I = \unit[1\times 10^{12}]{W / cm^2}$ (Fig.~\ref{fig:rdf_OO}a), the change in structure is dependent on the pump pulse frequency: at the frequencies that were shown to be most efficient at transferring energy to the water molecules, i.e., 16 and 19 THz \cite{mishra18}, the RDF is evened out more than at 7 and 30 THz, which indicates more structural changes are induced. At $I = \unit[5\times 10^{12}]{W /cm^2}$ (Fig.~\ref{fig:rdf_OO}b), the loss of order in the liquid is much more apparent. The RDF evens out, and the bulk water approaches a supercritical state of liquid density but gas-phase-like lack of order, regardless of the pump pulse frequency. Note that the employed PBE functional overestimates the degree of initial structural correlations compared to neutron diffraction data \cite{soper00, lin12}.

% Hydrogen bonds
We evaluate additional quantities at $t = \unit[250]{fs}$ directly from the AIMD trajectories, see Table~\ref{tab:aimd-quantities}. The fraction of hydrogen bonds remaining after the pulse is given. A hydrogen bond is here defined geometrically \cite{humphrey96} between two molecules where the oxygen-oxygen distance is less then $d_\mathrm{OO} = \unit[3.0]{\mathring A}$, and the angle between the covalently bound hydrogen and the acceptor oxygen is less then $\alpha = 20°$. For the lower $I = \unit[1\times 10^{12}]{W/cm^2}$, the remaining fraction of hydrogen bonds is frequency-dependent. With the stronger $I = \unit[5\times 10^{12}]{W/cm^2}$, the number of hydrogen bonds is depleted strongly regardless of the frequency. 
% saturation effects
Both the structural changes seen from the RDF as well as the depletion of hydrogen bonds are frequency-dependent only for the lower intensity. When taking into account the T-jumps (repeated from Ref.~\incite{mishra18} in Table~\ref{tab:aimd-quantities}), we note that, though the T-jump at the higher intensity and \unit[19]{THz} is about twice the T-jump at \unit[7 and 30]{THz}, the structural changes are comparable. We conclude that there is a saturation regime of structural changes with respect to the energy transferred by the THz pump pulse, and that it has been reached with the higher intensity. 

% FIGURE Broadening of the geometrical distributions
\begin{figure}
    \centering
    \includegraphics{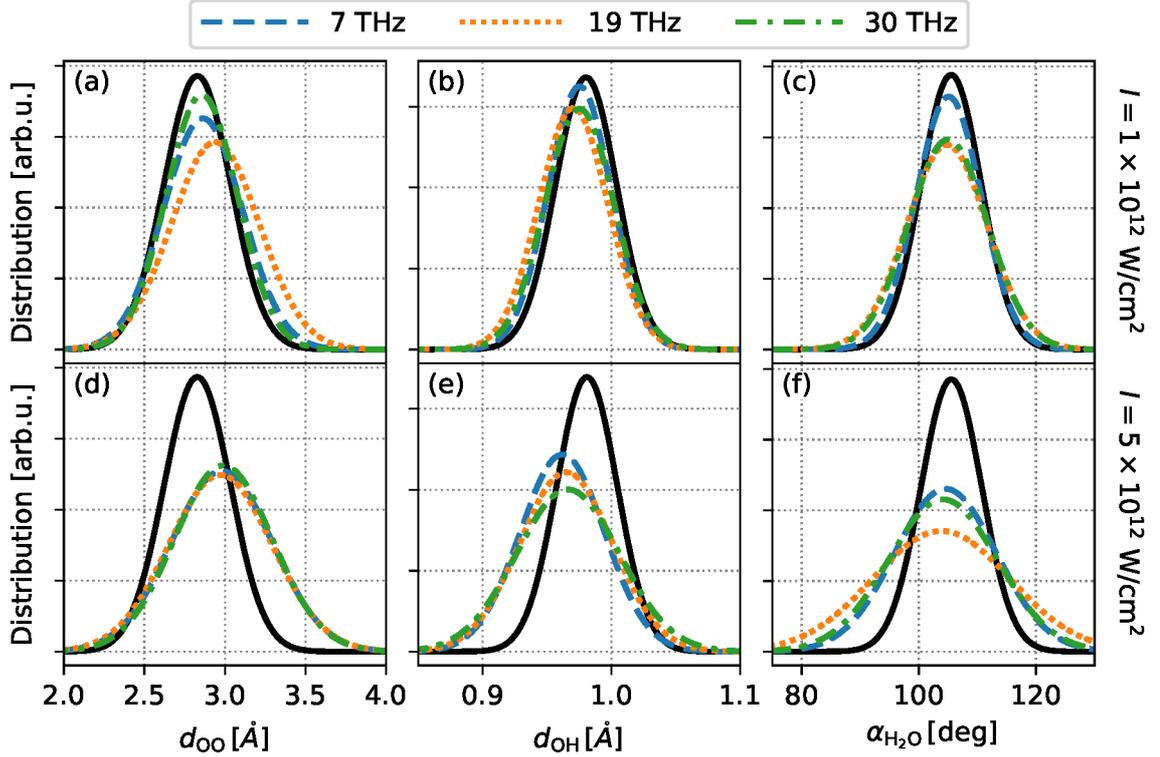}
    \caption{Distribution of various intra- and intermolecular distances and angles across trajectories, obtained after the THz pulse $(t = \unit[250]{fs})$: For $I = \unit[1\times 10^{12}]{W/cm^2}$ and $\nu =$~\unit[7, 19, and 30]{THz}, the distributions of \textit{(a)}~the distance between neighboring oxygen atoms $d_\chem{OO}$, \textit{(b)}~the covalent bond length $d_\chem{OH}$, and \textit{(c)}~the intramolecular angle $\alpha_\chem{H_2O}$ are displayed. The corresponding distributions for $I = \unit[5\times 10^{12}]{W/cm^2}$ are shown in panels $\textit{(d) -- (f)}$. For all quantities, the initial distribution obtained from unheated water $(t=\unit[-250]{fs})$ is included \textit{(solid black line)}. See Table~S1 in the SM for the values of the mean and width of the displayed distributions.}
    \label{fig:geometrical-distributions}
\end{figure}

% intramolecular, intermolecular
Overall, the THz pump pulse deposits a large amount of energy in the water, giving rise to a larger variability in structural arrangements of the water molecules. Comparing the unpumped and pumped water, the widths of distributions for different intra- and intermolecular distances and angles are increased. This is shown in Fig.~\ref{fig:geometrical-distributions} for the distance between neighboring oxygen atoms, the covalent bond length, and the intramolecular bending angle at lower intensity, $I = \unit[1\times 10^{12}]{W/cm^2}$~, panel \textit{(a)--(c)}, as well as at higher intensity, $I = \unit[5\times 10^{12}]{W/cm^2}$, panel \textit{(d)--(f)}. Moreover, small shifts in the means of these distributions can be observed. Both the broadening and the shifts increase with intensity. The values of the mean and width of these distributions can be found in Table~S1 of the supplemental material (SM).

\begin{figure}[h]
  \centering
  \includegraphics{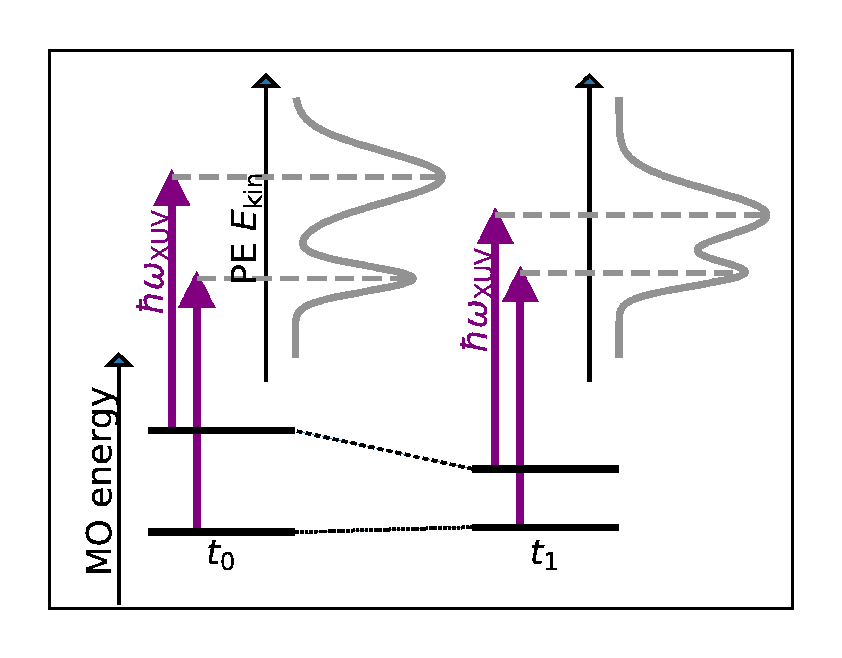}
  \caption{Schematic description of XUV photoelectron spectroscopy as a probe for chemical shifts. XUV photons with energy $\hbar\omega_\mathrm{XUV}$ create photoelectrons (PE) with kinetic energies given by the difference of the photon energy and the molecular orbital energy. Ultrafast structural changes induce chemical shifts \cite{link09a}. From time $t_0$ (left) to $t_1$ (right), the molecular orbitals are shifted in energy, which is directly reflected in the photoelectron spectrum. }
  \label{fig:xuvpes_sketch}
\end{figure}

Structural changes in water can induce chemical shifts of valence electrons, that then alter the photoelectron spectrum \cite{link09a}, see Fig.~\ref{fig:xuvpes_sketch}. In the following we investigate the photoelectron spectrum of T-jumped liquid water excited by THz pump pulses of varying intensities and frequencies and analyze changes in the valence photoelectron spectrum as an experimental signature of structural changes.

\section{Calculation of Photoelectron Spectra of Liquid Water}
\label{sect:photoionization-theory}

The valence orbitals of water can be probed using XUV photons \cite{banna86,winter04}. We consider a probe pulse of $\omega_\mathrm{XUV} = \unit[60]{eV}$ photon energy, and calculate both photoionization cross sections and photoelectron energies. We do not account for attenuation of the ejected photoelectrons in bulk water and note that, generally, photoelectrons probe the surface of liquids as attenuation lengths are on the order of few nanometers \cite{suzuki14}.

\subsection{\textsc{Xmolecule} Photoionization Cross Sections}

The photoionization cross sections of THz-pumped water are calculated using an extended version of the \textsc{Xmolecule} toolkit \cite{hao15,inhester16}. The outgoing photoelectron is described by atomic continuum wave functions, and its energy is obtained by Koopmans' theorem as $\varepsilon = \omega_\mathrm{XUV} + \varepsilon_a$, where $\varepsilon_a$ is the Hartree-Fock (HF) energy of the ionized molecular orbital (MO). MOs are represented in \textsc{Xmolecule} as linear combinations of atomic orbitals (LCAO) \cite{hao15}. Following the approximations of Ref.~\incite{gelius72}, the photoionization cross section for linearly polarized XUV light from MO $\phi_a$, averaged over the orientation of the molecules, is given by\cite{inhester16}
\begin{widetext}
\begin{equation}
    \begin{split}
    \sigma_a\left(\omega_\mathrm{XUV}\right)
    =
    \frac{4 \pi^2 \alpha \omega_\mathrm{XUV}}{3}
    \sum\limits_A^{\mathrm{atoms}} \sum\limits_{lm} \sum\limits_{d=x,y,z}
    \Bigg[
        \sum\limits_{\mu \mathrm{\,on A}} C_{\mu a}^2 |\braket{\chi_\mu | d | \chi_{\varepsilon l m}}|^2
        \\
        + 
        2 \sum\limits_{\mu < \nu\mathrm{\,on A}} C_{\mu a}C_{\nu a} \braket{\chi_\mu | d | \chi_{\varepsilon l m}}\braket{\chi_\nu | d | \chi_{\varepsilon l m}}
    \Bigg]
    ,
    \end{split}
    \label{eq:photionization-cross-section}
\end{equation}
\end{widetext}
where $\alpha$ is the fine-structure constant, $\chi_\mu$ a basis function in a minimal atomic orbital basis set, and $\chi_{\varepsilon l m }$ the wave function of the outgoing photoelectron with angular quantum numbers $l,m$ and energy $\varepsilon$. The transition dipole matrix elements $\braket{\chi_\mu| d | \chi_{\varepsilon l m}}$ are obtained from tabulated values from an atomic electronic structure calculation performed with \textsc{Xatom} \cite{jurek16}. Further details are given in previous work \cite{inhester16}. For calculations with the larger basis set, a projection of the MOs expanded in the larger basis set onto the minimal basis set is performed to obtain an LCAO description of the MOs with minimal basis functions. Accordingly, the corresponding LCAO coefficients for the minimal basis set are given by 
\begin{equation}
    \mathbf C^{(m)} = \mathbf S^{-1}\cdot \mathbf O \cdot \mathbf C^{(l)},
  \label{eq:projection}
\end{equation}
where $\mathbf C^{(l)}$ is the matrix of orbital coefficients in the larger basis set, $\mathbf{C}^{(m)}$ are the orbital coefficients in the minimal basis set, $\mathbf O$ is the overlap matrix between the different basis sets, and $\mathbf S^{-1}$ the inverse of the overlap matrix of the minimal basis set.

\subsection{Embedded Water Clusters}

As the investigation of liquid water requires sampling of many water cluster geometries, an efficient electronic structure method is imperative. The calculations are done at the HF level of theory employing the 6-31G basis set on water clusters containing 20 water molecules, which are embedded in point charges that model the surrounding water molecules. From a given trajectory snapshot, we select a $\chem{(H_2O)}_{20}$ water cluster around a central molecule corresponding to roughly two solvation shells around the central water molecule $(r \approx \unit[4.9]{\mathring A})$. At this cluster size, spectral features are converged with our electronic structure method, see Fig.~S14 of the SM. For water molecules beyond the second solvation shell, it is sufficient to consider their charge distribution rather than include them in the quantum electronic structure calculation \cite{yin17}. Thus, 108 surrounding molecules are substituted with point charges of $q_O = -0.82$ at the oxygen positions and $q_H = +0.41$ at each of the hydrogen positions, according to the charge values of the SPC 3-site water model \cite{berendsen81}. To sample different water geometries, photoelectron spectra are averaged from 43 $\chem{(H_2O)_{20}}$ clusters per trajectory snapshot. The dependence of photoelectron spectra on different water geometries is shown in Fig.~S15 of the SM. The calculated spectral transitions are convolved with Gaussians of $\unit[1.0]{eV}$ bandwidth to account for further broadening effects.

%\begin{table}
%    \caption{Peaks in liquid water, labeled with the molecular orbitals of gas-phase $\chem{(H_2O)}$ and their extension along the binding energy axis as obtained from Hartree-Fock electronic structure calculations.}
%  \label{tab:h2o-peak-ranges}
%  \begin{tabular}{ccc}
%    \toprule
%    Peak    &   From [eV]    &   To [eV] \\
%    \midrule
%    $2a_1$  &  42   &   30 \\
%    $1b_2$  &  24   &   19 \\
%    $3a_1$  &  19   &   14 \\
%    $1b_1$  &  14   &   9 \\
%    \bottomrule
%  \end{tabular}
%\end{table}

% ================================+
% RESULTS SECTION
\section{Photoelectron Spectroscopy of THz-Pumped Water}
\label{sect:results}

\subsection{Photoelectron Spectrum of Unheated Water}
\label{sect:results-unheated}

\begin{figure}[h]
    \centering
    \includegraphics{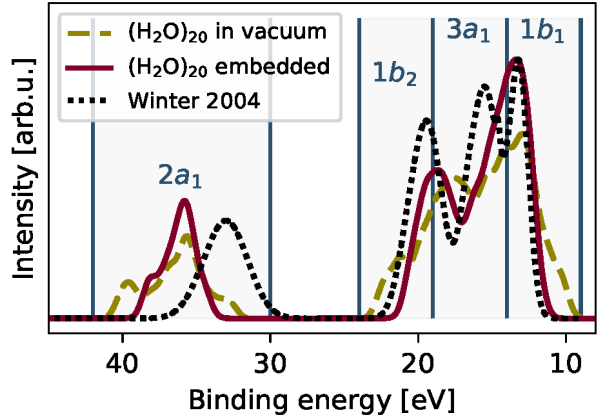}
    \caption{Photoelectron spectrum of unheated liquid water from $\chem{(H_2O)_{20}}$ water clusters in vacuum \textit{(yellow dashed line)}, averaged over 43 geometries from ten trajectory snapshots, and the same clusters embedded in 108 additional water molecules modeled as point charges $(q_O = -0.82, q_H = +0.41)$ \textit{(velvet solid line)}. Experimental data \cite{winter04} is included for comparison \textit{(black dotted line)}, shifted by \unit[2]{eV} towards lower binding energies and rescaled to the outer-valence peak height obtained from embedded $\chem{(H_2O)_{20}}$ clusters. The arising peaks are labeled with the molecular orbitals of a single $\chem{H_2O}$ for later reference. Their extension along the binding energy axis, as obtained from Hartree-Fock electronic structure calculations, is \unit[42 to  30]{eV} $(2a_1)$, \unit[24 to  19]{eV} $(1b_2)$, \unit[19 to  14]{eV} $(3a_1)$, and \unit[14 to  9]{eV} $(1b_1)$.  
  }
    \label{fig:qm-mm-comparison}
\end{figure}

The simulated photoelectron spectrum of liquid water before the action of the THz pulse is shown in Fig.~\ref{fig:qm-mm-comparison}. The photoelectron spectrum obtained from the clusters in vacuum is compared to the spectrum from embedded clusters. From higher to lower binding energies, the arising peaks are labeled with the molecular orbitals of a single $\chem{H_2O}$. Their extension along the binding energy axis, as obtained from Hartree-Fock electronic structure calculations, is \unit[42 to  30]{eV} $(2a_1)$, \unit[24 to  19]{eV} $(1b_2)$, \unit[19 to  14]{eV} $(3a_1)$, and \unit[14 to  9]{eV} $(1b_1)$. The low-lying $1a_1$ orbital is formed mostly from the oxygen $1s$ orbital and is beyond the energy of XUV photons. The embedding leads to a more compact photoelectron spectrum, as surface effects are diminished. For comparison, an experimental XUV photoelectron spectrum of liquid water is included \cite{winter04}. Most of the relevant spectral features, peak positions, widths, and relative heights, are well reproduced within our theory. The gap between the outer- and inner-valence peaks is overestimated due to the lack of relaxation mechanisms in the employed electronic structure model. While the $1b_2$ peak is clearly discernible, the $3a_1$ and $1b_1$ peak are less separated than seen in experimental data. This is consistent with electronic structure calculations on gas phase $\chem{H_2O}$, that show the gap between these peaks to be \unit[1.43]{eV}, underestimating the experimental gap of \unit[2.24]{eV}\cite{winter04}. In the following, we concentrate on relative changes in the photoelectron spectrum induced by THz-pumping, and the systematic errors introduced by the electronic structure method mostly cancel.

\subsection{Photoelectron Spectrum of T-jumped Water}
\label{sect:results-single-pulse}
\begin{figure*}[h]
  \centering
  \includegraphics{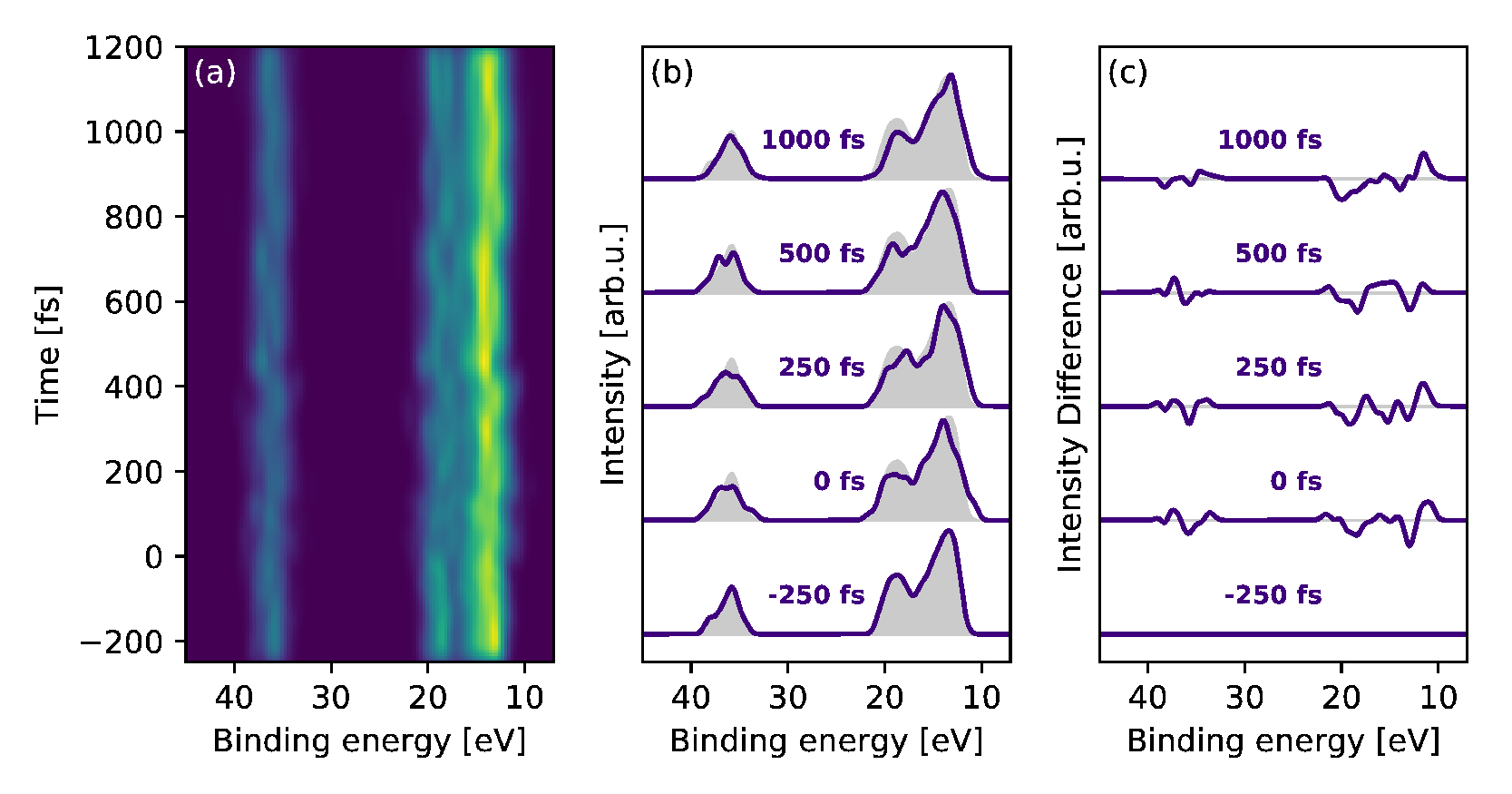}
  \caption{Time evolution of the photoelectron spectrum for incident XUV photon energy $\unit[60]{eV}$, THz pump pulse parameters $I = \unit[5\times 10^{12}]{W/cm^2}, \nu = \unit[19]{THz}$, and $\Delta \tau_\mathrm{FWHM} = \unit[150]{fs}$. Here, spectra from 43 geometries of $\chem{(H_2 O)_{20}}$ from seven trajectory snapshots, embedded in an additional 108 molecules that are modeled as point charges, were averaged. \textit{(a)} Time evolution of the photoelectron spectrum to $\unit[1200]{fs}$, convolved in time with the expected XUV pulse duration of $\unit[50]{fs}$ for comparison with future experimental data. \textit{(b)} Photoelectron spectra for distinct time steps, where $t = \unit[0]{fs}$ refers to the maximum of the THz pulse envelope. The initial photoelectron spectrum of the unheated water is shown as the shaded area. \textit{(c)} Difference spectra for distinct time steps with respected to the photoelectron spectrum of unheated water at $\unit[t=-250]{fs}$. }
  \label{fig:complete-picture}
\end{figure*}
\begin{figure*}
    \centering
    \includegraphics{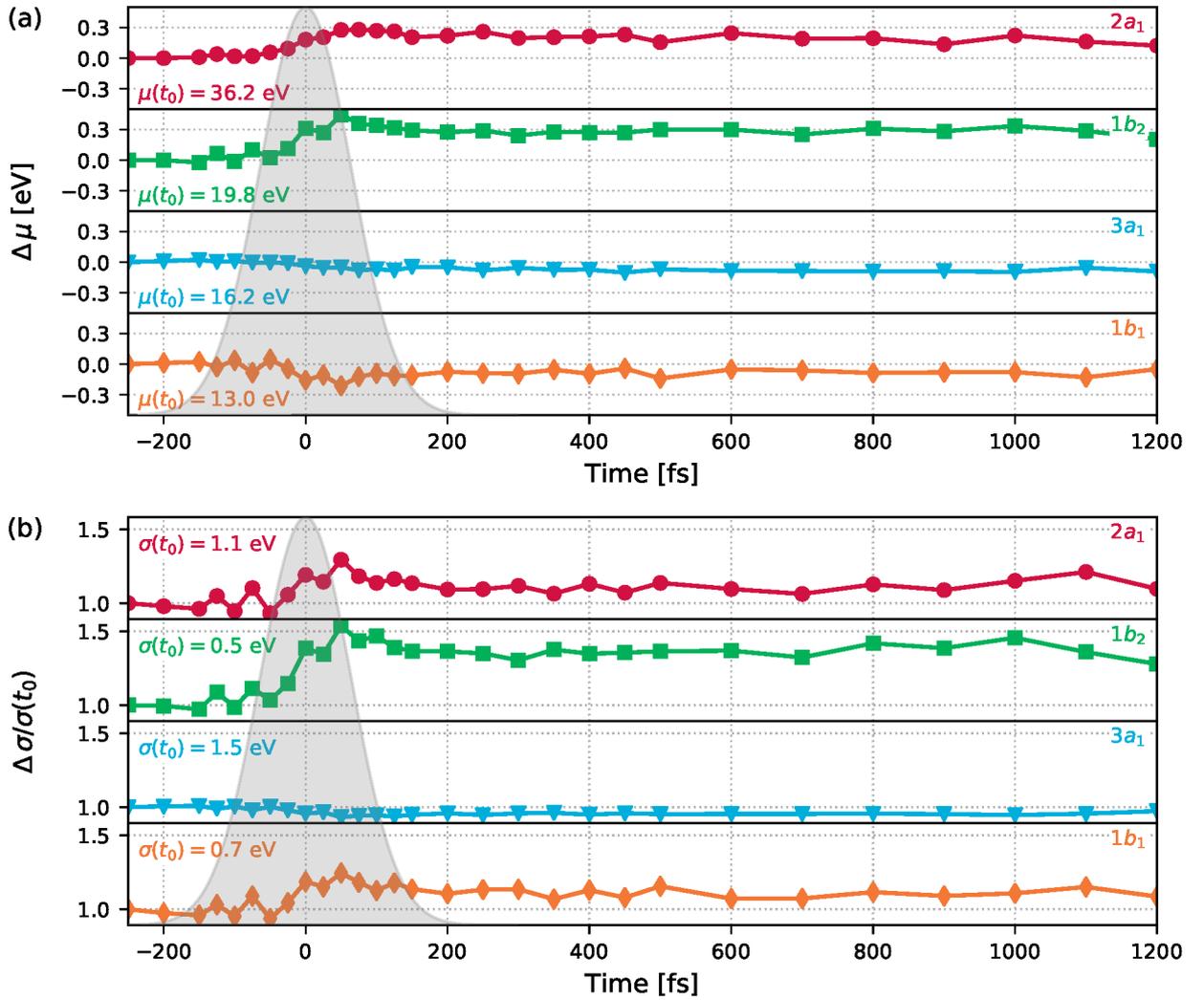}
    \caption{
        Time evolution of the means \textit{(a)} and widths \textit{(b)} of the peaks in the photoelectron spectrum induced by the same pulse as in Fig.~\ref{fig:complete-picture}. The change of peak means and the relative change in peak widths is plotted, with the initial values at $t_0 = \unit[-250]{fs}$ indicated in the respective panels. The THz pump pulse intensity is included as the shaded grey area. From top to bottom, the peaks shown correspond to $2a_1, 1b_2, 3a_1,$ and $1b_1$, cf.~Fig.~\ref{fig:qm-mm-comparison}.
    }
    \label{fig:mean_width}
\end{figure*}
We now calculate relative changes induced through THz-pumping in the photoelectron spectra with respect to the initial spectrum shown in Fig.~\ref{fig:qm-mm-comparison}. This allows us to quantify the impact of sub-picosecond THz-induced T-jumps on photoelectron spectra of bulk water. We focus on pulse parameters that induce a strong T-jump of about $\unit[1200]{K}$ \cite{mishra18}, i.e., $I = \unit[5\times 10^{12}]{W/cm^2}$ and $\nu = \unit[19]{THz}$. Figure~\ref{fig:complete-picture} shows the time evolution of the photoelectron spectrum of embedded $\chem{(H_2O)_{20}}$ clusters, averaged over 43 water cluster geometries per trajectory snapshot. Changes in the photoelectron spectrum are highlighted by comparing the photoelectron spectrum along the trajectories with the initial, unperturbed one (Fig.~\ref{fig:complete-picture}b, shaded grey area), and by including the difference spectrum with respect to the initial photoelectron spectrum (Fig.~\ref{fig:complete-picture}c).

To quantify these changes, we identify the peaks labeled by the MOs of a single $\chem{H_2O}$ 
at the intervals described above (cf.~Fig.~\ref{fig:qm-mm-comparison}). For each interval, we calculate the mean and width as the first cumulant and square root of the second cumulant of the spectrum, respectively. The resulting changes in the photoelectron spectra are shown in Fig.~\ref{fig:mean_width}. The inner-valence shell $2a_1$ peak, as well as the $1b_2$ peak, shift towards higher binding energies by up to $\Delta \mu(t) = \mu(t) - \mu(t_0) \approx \unit[0.4]{eV}$. These two peaks are formed by MOs with strong binding character and are thus affected by the heating of intramolecular degrees of freedom. The $3a_1$ peak, formed by MOs that contribute to OH-bonding\cite{guo10} shifts slightly to lower binding energies by about \unit[0.1]{eV}. Finally, the $1b_1$ peak, characterized by MOs from the oxygen lone pairs, stays unaffected. The relative increases in peak widths, $\Delta \sigma(t) = \left(\sigma(t) - \sigma(t_0)\right)/\sigma(t_0)$, amount to about $1.2$ for the $2a_1$ and $1.5$ for the $1b_2$ peak. The $3a_1$ peak is constant in width, and the $1b_1$ peak again extends slightly by $1.1$. Taken together, the three outer-valence peaks are broadened. After the pulse has ended at $t = \unit[250]{fs}$, the centers and widths of the peaks in the photoelectron spectrum only show small changes around the then-achieved values. The details of the transient changes occuring during the THz pulse are discussed in the SM.

The temporal variations in the photoelectron spectra indicate changes in the chemical environment and are associated with the transformation from liquid bulk water to supercritical water-like structures, i.e., gas with liquid density. Intramolecular vibrations are increasingly heated \cite{mishra18}, and hydrogen bonds are broken. The sampled water cluster geometries show greater variation in intra- and intermolecular degrees of freedom than in the unheated water, which explains the increasing peak widths. 
%The changes during the THz pump pulse will be analyzed in more detail in Sec.~\ref{sect:results-different-pulses}.  
The non-uniform initial orientation of molecular dipole moments (cf.~Sec.~\ref{sect:aimd-trajectories}) is not expected to affect these conclusions, as the deposition of large amounts of energy through the THz pulse will quickly decrease any artificial local order.
From the analysis of trajectories with different THz frequencies, we observe that the changes in photoelectron spectrum peak centers that remain after the pulse $(t = \unit[250]{fs})$ are not frequency-dependent within the statistical error (see Fig.~S1 in the SM). However, the T-jump at $I = \unit[5\times 10^{12}]{W/cm^2}$ was found to be about $\unit[720]{K}$ for $\unit[7]{THz}$ and $\unit[560]{K}$ for $\unit[30]{THz}$, respectively \cite{mishra18}. This indicates that the energy transferred by the high-intensity pulse exceeds a saturation threshold, in the sense that the changes in the photoelectron spectra do not reflect the amount of the T-jump. 

\begin{figure}
    \centering
    \includegraphics{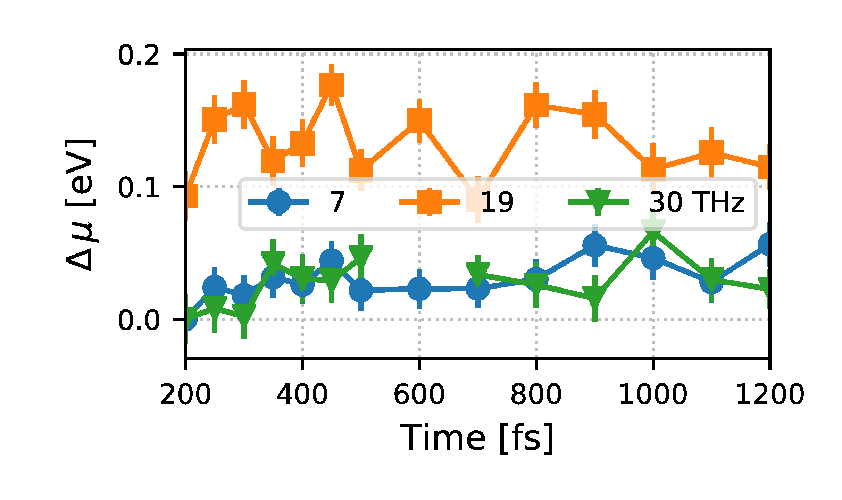}
    \caption{Time evolution of the $2a_1$ peak in the photoelectron spectrum induced by a THz pulse of $I=\unit[1\times 10^{12}]{W/cm^2}$ and $\nu = $\unit[7, 19, and 30]{THz}. The changes during the pulse duration are omitted for clarity.}
    \label{fig:thermometer_1-12_2a1}
\end{figure}

For a lower pump pulse intensity of $I = \unit[1\times 10^{12}]{W/cm^2}$, the changes in peak positions are qualitatively the same, but now with a maximum variation of only $\unit[0.2]{eV}$. A full analysis of the photoelectron spectrum at lower intensity can be found in Fig.~S2 of the supplemental material (SM). Here, we focus on the $2a_1$ peak, that was most affected by the heating process. In Fig.~\ref{fig:thermometer_1-12_2a1}, the time evolution of this peak position is shown for $I =  \unit[1\times 10^{12}]{W/cm^2}$ and pump pulse frequencies of 7, 19, and 30 THz. The frequency dependence that was discussed in Sec.~\ref{sect:aimd-quantities} is visible, since pumping with $\nu = \unit[19]{THz}$ induces up to twice the change compared to pumping with $\nu = \unit[7, 30]{THz}$. The corresponding T-jumps are about $\unit[210]{K}$ for 19 THz, $\unit[80]{K}$ for 7 THz, and $\unit[130]{K}$ for 30 THz, see Table~\ref{tab:aimd-quantities}. They are thus well below the $\unit[1200]{K}$ that could be achieved by high-intensity THz pulses. 

\section{Outlook and Conclusion}
\label{sect:outlook}

We investigated the possibility of detecting ultrafast THz-induced T-jumps in liquid water by employing XUV photoelectron spectroscopy. The structural changes induced by sub-picosecond, high-intensity THz pulses of various frequencies were analyzed by calculating RDFs, building on AIMD trajectories that were the subject of previous work \cite{mishra18}. We found that, for a lower intensity, the loss of order is frequency-dependent. It is highest at the frequencies that are associated with the largest energy transfer. At higher intensity, a supercritical phase of liquid density but gas-phase disorder is formed regardless of the frequency. We showed the impact of the induced disorder in nuclear geometry on the electronic structure by simulating XUV photoelectron spectra. We found that for high pump-pulse intensity, the peak centers shift by up to \unit[0.4]{eV}, and the peak widths are broadened. 

This work relies on the only AIMD trajectories of THz-pumped, T-jumped liquid water that are currently available at the high intensity and different frequencies we are interested in. Due to the large number of pump pulse intensities and frequencies considered in Ref.~\incite{mishra18}, the simulation protocol was restricted. The PBE functional was chosen based on the good description of electronic polarizability \cite{mishra13, mishra18}, however this resulted in overly structured water \cite{lin12, soper00} and subpar equilibration in the generation of initial configurations. The deficiencies could be overcome by using alternative density functionals \cite{lin12} or by scaling to a higher initial temperature \cite{heyden10}. However, such improvement is beyond the scope of this work. The large amount of energy transferred by the THz pulse is expected to overcome any deficient initial equilibration. The electronic structure calculations were performed at the Hartree-Fock level of electronic structure theory, since the time-dependent investigation of XUV photoelectron spectra of liquid water called for an efficient electronic structure method. Since we focus on relative changes of the photoelectron spectra, the systematic errors in the electronic structure calculations are expected to mostly cancel. 

During the temporal overlap of the THz pump and the XUV probe pulse, the photoelectrons generated by the probe pulse are distorted by the laser field of the pump pulse. This leads to changes in the photoelectron spectrum known as THz streaking \cite{fruhling09}. The streaking effect can approximately be calculated analytically \cite{fruhling11} and is widely employed as a per-shot analysis tool for femtosecond pulses, as it allows inferring time-domain properties of the pulse by identifying them with changes in the photoelectron spectrum. Investigating the THz streaking effects in connection with XUV photoelectron spectroscopy of THz-pumped water is left for future work.

Regarding future experiments, we note that the reported changes in the photoelectron spectrum are small and close to the spectral resolution currently obtained with good monochromators \cite{poletto10, gerasimova11}. Photoelectron spectroscopy of liquid-phase systems poses additional technical and instrumentation challenges, but has been demonstrated \cite{winter06, faubel12, kolbeck12}. Since the overall changes in the photoelectron spectrum do not reflect the frequency-dependency of the T-jump at high THz pulse intensity, photoelectron spectroscopy on its own should not be seen as a thermometer for ultrafast heating of liquid water. It provides an interesting complementary diagnostics tool in combination with x-ray diffraction studies and shows how the electronic system adapts to the structural changes induced by the pump pulse. 

%================================================================================ 
% SUPPORTING INFORMATION
%================================================================================ 

\section*{Supplementary Material}
Changes in photoelectron spectra induced by pump pulses with $I = \unit[5\times 10^{12}]{W/cm^2}, \nu = $\unit[7 and 30]{THz} and $I = \unit[1\times 10^{12}]{W/cm^2}, \nu = \unit[19]{THz}$; Transient changes during the pump pulse; Analysis of transient and final geometrical changes and hydrogen-bond breaking; Electronic structure method.

%================================================================================ 
% ACKNOWLEDGMENTS
%================================================================================ 

\begin{acknowledgments}
We thank Pankaj Kumar Mishra for fruitful discussions. This work has been supported by the excellence cluster \textit{The Hamburg Centre for Ultrafast Imaging - Structure, Dynamics and Control of Matter at the Atomic Scale} of the Deutsche Forschungsgemeinschaft.
\end{acknowledgments}

%================================================================================ 
% LITERATURE
%================================================================================ 
%   \bibliographystyle{jpc}
%   \bibliography{2017_DISS06_THzWater}

\end{document}